\documentclass[final,3p,times,twocolumn]{elsarticle}

\usepackage{amssymb}
\usepackage{epsfig}

\journal{Physica C - special issue on Fe arsenides}

\begin{document}

\begin{frontmatter}



\title{High-field phase-diagram of Fe arsenide superconductors}


\author{Y. J. Jo \fnref{NHMFL}}\author{J. Jaroszynski \fnref{NHMFL}}\author{A. Yamamoto \fnref{NHMFL}}\author{A. Gurevich \fnref{NHMFL}}\author{S. C. Riggs \fnref{NHMFL}}\author{G. S. Boebinger \fnref{NHMFL}}\author{D. Larbalastier \fnref{NHMFL}}\author{H. H. Wen \fnref{beijing}}\author{N.\ D.\ Zhigadlo \fnref{ETH}}\author{S.\ Katrych \fnref{ETH}}
\author{Z.\ Bukowski \fnref{ETH}}
\author{J.\ Karpinski \fnref{ETH}}
\author{R. H. Liu \fnref{Hefei}}
\author{H. Chen \fnref{Hefei}}
\author{X. H. Chen \fnref{Hefei}}
\author{L. Balicas \fnref{NHMFL}}
\ead{balicas@magnet.fsu.edu}

\address[NHMFL]{National High Magnetic Field Laboratory, Florida
State University, Tallahassee-FL 32310, USA}

\address[beijing]{Institute of Physics, Chinese Academy of Sciences, Beijing 100190, People's Republic of China}

\address[ETH]{Laboratory for Solid State Physics, ETH
Z\"{u}rich, CH-8093 Z\"{u}rich, Switzerland}

\address[Hefei]{Hefei National Laboratory for Physical Science a
Microscale and Department of Physics, University of Science and
Technology of China, Hefei, Anhui 230026, People's Republic of
China}

\begin{abstract}
Here, we report an overview of the phase diagram of single layered
and double layered Fe arsenide superconductors at high magnetic
fields. Our systematic magnetotransport
measurements of polycrystalline SmFeAsO$_{1-x}$F$_x$ at different
doping levels confirm the upward curvature of the upper
critical magnetic field $H_{c2}(T)$ as a function of temperature
$T$ defining the phase boundary between the superconducting
and metallic states for crystallites with the ab planes oriented
nearly perpendicular to the magnetic field. We further show from
measurements on single crystals that this feature, which was
interpreted in terms of the existence of two superconducting gaps,
is ubiquitous among both series of single and double layered
compounds. In all compounds explored by us the zero temperature
upper critical field $H_{c2}(0)$, estimated either through the
Ginzburg-Landau or the Werthamer-Helfand-Hohenberg single gap
theories, strongly surpasses the weak coupling Pauli paramagnetic
limiting field. This clearly indicates the strong coupling nature of
the superconducting state and the importance of magnetic correlations for
these materials. Our measurements indicate that the superconducting
anisotropy, as estimated through the ratio of the effective masses
$\gamma = (m_c/m_{ab})^{1/2}$ for carriers moving along the c-axis
and the ab planes, respectively, is relatively modest as compared to
the high-$T_c$ cuprates, but it is temperature, field and even doping
dependent. Finally, our preliminary estimations of the
irreversibility field $H_m(T)$, separating the vortex-solid from the
vortex-liquid phase in the single layered compounds, indicates that it is well described by
the melting of a vortex lattice in a moderately anisotropic uniaxial superconductor.
\end{abstract}


\end{frontmatter}


\section{Introduction}
\label{}

The iron-based oxypnictides represent a novel class of
superconductors which, with the exception of the cuprates, have the
highest known superconducting transition temperature $T_c$
\cite{materials, chennature}. Several series of these compounds have
been synthesized in the last year, but throughout this manuscript we
will mostly focus on the properties of the single-layered
oxypnictide \emph{Ln}FeAsO$_{1-x}$F$_x$ (\emph{Ln} is a lanthanide),
or the so-called 1111 compounds, and on the \emph{AE}Fe$_2$As$_2$
(\emph{AE} is an alkali-earth metal), or the so-called 122
compounds.

Both electric transport measurements and electronic band structure
calculations suggest that undoped oxypnictides are semimetals
\cite{bandstructure}. There is an approximate nesting between the
hole Fermi surface (FS) centered at the $\Gamma$ point and the
electron FS centered at the \emph{M} point, which may lead
to a spin-density wave (SDW) like instability state observed at low
temperatures \cite{mazin,dong}. According to neutron diffraction
studies, the magnetic structure of undoped pnictides is composed of
antiferromagnetically coupled ferromagnetic chains \cite{dai}.
Superconductivity in the 1111 or electron doped compounds would occur when part of the Fe$^{2+}$ ions are
replaced by Fe$^+$, which is expected to suppress the
antiferromagnetic instability.

Several superconducting pairing mechanisms based on the multi-band
nature of these compounds have been proposed. Dai et al.
\cite{daitriplet} suggested a spin-triplet pairing mechanism with
even parity due to ferromagnetic spin fluctuations between electrons
in different orbitals. Lee and Wen \cite{palee} argued that the
strong Hund's rule ferromagnetic interaction in Fe pnictides can
lead to a pairing instability in the spin-triplet \emph{p}-wave channel in
the weak coupling limit, so that the superconducting gap would have
nodal points on the two-dimensional Fermi surfaces. While Lee et al.
suggested that because of the frustrating pairing interactions among
the electron and the hole fermi-surface pockets, a $s + id$ pairing
state with broken time reversal symmetry could be favored
\cite{lee}. Perhaps, the model that is currently more widely
accepted is the so-called extended $s^\pm$-wave model, which
predicts a $\pi$ shift between the order parameters on the hole and
the electron Fermi surface sheets \cite{mazin}. In this scenario,
the unconventional pairing mechanism is mediated by
antiferromagnetic spin fluctuations. In fact, inelastic neutron
scattering in the Ba$_{0.6}$K$_{0.4}$Fe$_2$As$_2$ compound reveals
the emergence of a localized resonant magnetic excitation below the superconducting
transition temperature \cite{n}. This type of excitation is expected
for a superconducting order parameter which has opposite signs in
different parts of the Fermi surface as in the $s^\pm$ scenario.
A general overview of the different pairing scenarios in oxypnictides is given by Mazin and Schmalian in this volume
\cite{mazz}.

It is  interesting to mention a recent photoemission report \cite{zabolotnyy} claiming
the existence of an underlying electronic $(\pi, \pi)$ order in
Ba$_{1-x}$K$_x$Fe$_2$As$_2$, previously seen in the cuprates
\cite{electrondoped} and claimed to be perhaps at the origin of
the observed small Fermi surface pockets, as also seen in underdoped
YBa$_2$Cu$_3$O$_{7 + \delta}$ \cite{taillefer}. Thus the coexistence and
non-trivial interplay of different order parameters seems to be an essential
ingredient for high temperature superconductivity, in either the
cuprates or the oxypnictides.

As for the existence of multiple superconducting gaps
owing to the multi-band nature of the Fe arsenides the situation is
still somewhat inconclusive. Several point contact spectroscopy
studies yield contradictory results, with evidence for both single
\cite{chien} and multi-gap superconductivity \cite{multiplegaps}.
The latter does not invalidate the multiband pairing scenario but
rather indicates that the gaps of approximately equal magnitudes
may reside on different disconnected sheets of the Fermi surface. At the
same time, detailed local magnetization measurements via Hall probe
magnetometry \cite{localmagnetization}, photoemission \cite{ding,richards} and neutron \cite{n}
measurements are consistent with a two-gap superconducting scenario for the 122
compounds.

The exact shape of the $H-T$ superconducting phase diagram can help
distinguish between different theoretical scenarios. Moreover,
magnetotransport measurements under strong magnetic field can also
determine the temperature dependence of the irreversibility field,
which is one of the key parameters quantifying the potential of the
oxypnictides for future power applications. However, it is already
clear \cite{hunte,jan} that these compounds are characterized by tremendously large
upper critical fields, requiring very high magnetic field techniques
to explore the overall complexity of their superconducting phase
diagram. This is in turn a very encouraging result for applied
superconductivity.

There have been already several reports on the physical properties
of the Fe arsenides at very high magnetic fields. Two relevant
reports, provide i) evidence for two-gap superconductivity to
explain the upward curvature of the upper critical field as a
function of temperature for fields along the c-axis direction in
1111 compounds \cite{hunte,jan}, and ii) an evidence that the
$H_{c2}(T)$ of the 122 compounds is weakly anisotropic despite the
two-dimensional character of their Fermi surface, thus indicating
that a strong anisotropy may not be instrumental for the high-$T_c$
superconductivity in the Fe arsenides \cite{singleton}.

One of the greatest challenges at the moment is to achieve the
synthesis of high quality single crystals, and to explore their
phase diagrams at very high fields not only by traditional
magnetotranport measurements but also by thermodynamic means.
Here, we provide a brief overview of our initial efforts in this direction.

\section{Experimental}

Polycrystalline samples with nominal composition
SmFeAsO$_{1-x}$F$_x$ were synthesized in Heifei by conventional
solid state reaction using high-purity SmAs, SmF$_3$, Fe and
Fe$_2$O$_3$ as starting materials. SmAs was obtained by reacting Sm
chips and As pieces at 600 $^{\circ}$C for 3 h and then 900
$^{\circ}$C for 5 h. The raw materials were thoroughly ground and
pressed into pellets. The pellets were wrapped in Ta foil, sealed in
an evacuated quartz tube, and finally annealed at either 1,160
$^{\circ}$C or 1,200 $^{\circ}$C for 40 h. X-ray diffraction (XRD)
pattern for a sample annealed at 1,160 $^{\circ}$C did reveal trace
amounts of the impurity phase SmOF \cite{chennature}.

For the growth of SmFeAsO$_{1-x}$F$_x$ single crystals at ETH
\cite{zhigadlo}, FeAs, Fe$_2$O$_3$, Fe and SmF$_3$ powders were used
as starting materials and NaCl/KCl was used as flux. The precursor
to flux ratio varied between 1:1 and 1:3. Pellets containing
precursor and flux were placed in a BN crucible inside a
pyrophyllite cube with a graphite heater. Six tungsten carbide
anvils generated pressure on the whole assembly (3 GPa was applied
at room temperature). While keeping pressure constant, the
temperature was ramped up within 1 h to the maximum value of
1350-1450 $^{\circ}$C, maintained for 4-10 h and decreased in 5-24 h
to room temperature for the crystal growth. Then pressure was
released, the sample removed and in the case of single crystal
growth NaCl/KCl flux was dissolved in water. Below, in our
discussion concerning the irreversibility line in the 1111
compounds, we include some measurements previously reported by us in
Ref. \cite{jan} in NdFeAsO$_{0.7}$F$_{0.3}$ single crystals (see
Ref. \cite{wenNd} for the details concerning the sample synthesis)
which were originally used to extract the boundary between metallic
and superconducting states at high fields.

The results reported here on 122 compounds were measured on single crystals of BaFe$_{2-x}$Co$_x$As$_2$ and Ba$_{1-x}$K$_x$Fe$_2$As$_2$ grown at Hefei by
the self-flux method. In order to avoid contamination from
incorporation of other elements into the crystals, FeAs was chosen
as self-flux. FeAs and CoAs powder was mixed together, then roughly
grounded. The Ba pieces were added into the mixture. The total
proportion of Ba:(2-xFeAs+xCoAs) is 1:4. For more details, see Ref.
\cite{xfwang}. A similar procedure was used to synthesize single
crystals of Ba$_{1-x}$K$_x$Fe$_{2}$As$_2$.

\begin{figure}[htb]
\begin{center}
\epsfig{file= 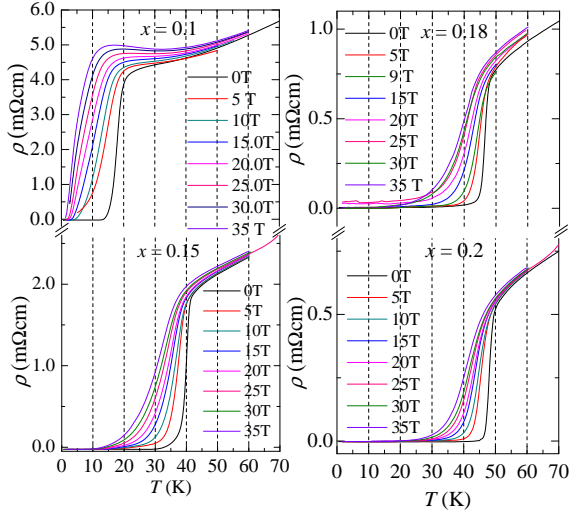, width = 7.4 cm} \caption {(color
online) Resistivity of SmFeAsO$_{1-x}$F$_x$ as a function of temperature for several magnetic field values and for four doping levels
$x = 0.1, 0.15, 0.18$ and 0.2.   }
\end{center}
\end{figure}

For electrical transport measurements, polycrystalline samples were
cut into bar shaped pieces. Contacts for the standard four probe measurements
were made by attaching gold wires with silver epoxy. Contacts in single crystals were
made by the focused-ion-beam technique as described in Ref.
\cite{wenNd}. For the magnetic torque measurements a
SmFeAsO$_{0.8}$F$_{0.2}$ single-crystal was attached to the tip of a
piezo-resistive micro-cantilever which was itself placed in a
rotator inserted into a vacuum can. Changes in the resistance of the
micro-cantilever associated with its deflection and thus a finite
magnetic torque $\tau$ was measured via a Wheatstone resistance
bridge. The ensemble was placed into a $^4$He cryostat coupled to a
resistive 35 T dc magnet. For the transport measurements we used a
combination of pulsed and continuous fields including the 45 T
hybrid magnet of the National High Magnetic Field Lab in
Tallahassee.

\section{Results and discussion}

\subsection{Polycrystalline material}
\begin{figure}[htb]
\begin{center}
\epsfig{file= 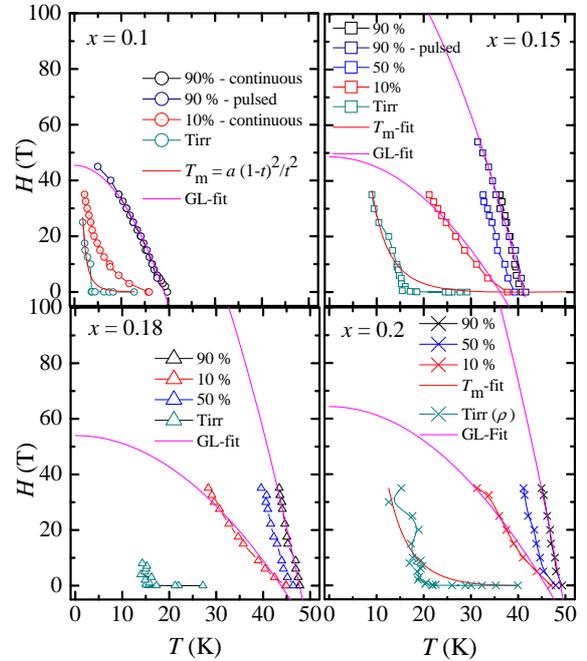, width = 7.4 cm} \caption {(color
online) The superconducting phase diagram of polycrystalline SmFeAsO$_{1-x}$F$_x$ for the four doping levels
shown in Fig. 1. Black and dark blue points represent the position of the onset (90\%), blue points represent middle point (50 \%) while red points represent the foot (10 \%) of the resistive transition. Magenta lines are fits to the expression $H_c=H_{c2}(0)[1-t^2]$ while red lines are attempts to describe onset of the zero resistance state (green markers) through an anisotropic vortex melting line, $T_m=a \frac{(1-t)^2}{t^2}$, where $t = T/T_c$.}
\end{center}
\end{figure}
\begin{figure}[htb]
\begin{center}
\epsfig{file= 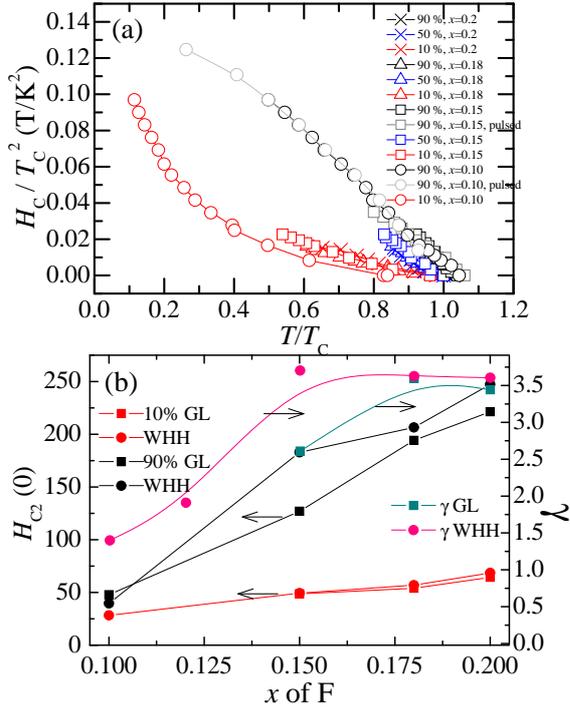, width = 7.4 cm} \caption {(color
online) Top panel: Renormalized superconducting phase diagram of the SmFeAsO$_{1-x}$F$_x$ series. Despite the known existence of sub-phases in polycrystalline material,
the observed scaling indicates that one is dealing with a material that is effectively in the clean limit. Bottom Panel: Estimations of the upper critical field at zero temperature $H_{c2}(0)$ from
the WHH formulae (solid circles) and from the fit to $H_c=H_{c2}(0)[1-t^2]$ (solid squares). Red and black markers correspond to fits to points describing the $H-T$ position of respectively the 10 \% ($\sim H_{c2}^{c}$) and 90 \% ($\sim H_{c2}^{ab}$) of the resistive transition. In the same graph we plot the anisotropy $\gamma = H_{c2}^{ab}/ H_{c2}^{c}$ resulting from each fit. $\gamma$ is seen to increase with F content saturating at a value of $\simeq 3.5$.}
\end{center}
\end{figure}

In Fig. 1 we show typical resistivity curves for polycrystalline
SmFeAsO$_{1-x}$F$_x$ as a function of temperature
under several values of magnetic field and for four nominal doping
levels, $x=0.1$, 0.15, 0.18, and 0.2, \cite{liuprl} respectively. The first
striking feature is that one does not see a very pronounced
broadening of the superconducting transition under strong fields, in
marked contrast with the thermally-activated broadening seen in the
cuprates \cite{ando}. Although, both polycrystalline and
single crystalline SmFeAsO$_{1-x}$F$_x$ samples exhibit clear
signatures for thermally-activated flux-flow \cite{jan} indicating
the existence of a vortex-liquid state over a broad range of
temperatures and magnetic fields. The broad peak seen in the
resistivity above the superconducting state for $x=0.1$ results from a remarkable
magnetoresistive effect observed only in the under-doped compounds, i.e. $x < 0.15$, a behavior previously reported by
us in Ref. \cite{riggs} where we drew a comparison with the underdoped cuprates.
In this case, the onset of the resistive transition under field was defined as 90 \% of the
value of the resistivity maximum in the normal state just above the transition.

The conventional way of analyzing the resistive transition in
a polycrystalline layered material like SmFeAsO$_{1-x}$F$_x$ is to
assume that the onset of the superconducting transition is dominated
by those crystallites with the ab planes oriented nearly along the magnetic
field direction. In turn, the bottom of the resistive transition (at
which the resistivity $\rho(T,H)$ drops below $10\%$ of its normal
state value $\rho_n$) would be dominated by either those
crystallites with the ab plane oriented perpendicular to the magnetic
field or by the melting of the vortex lattice, as has been observed in
the cuprates. Thus, by measuring the temperatures of the onset and
the foot of the resistive transition under field, one can infer the
the temperature dependencies of the upper critical fields $H_{c2}$
 for both field orientations. But a word of caution is needed here. For polycrystalline material the width of the superconducting transition under field is affected by several factors including vortex fluctuations  and the coupling strength  between the grains.  Thus, the foot of the transition in polycrystalline material may not reflect the actual behavior of $H_{c2}^c$ for fields applied along the inter-plane direction if the Ginzburg parameter is of the order of $10^{-2}$ like in YBCO \cite{jan}. In 122 compounds and in the La based 1111 compounds the fluctuation Ginzburg parameter Gi is of the order of $10^{-4}$ indicating that vortex fluctuations are not that important, and the foot of the resistive transition does reflect the true behavior of $H_{c2}^c(T)$ \cite{hunte}. The situation becomes more complicated in Nd and Sm 1111 compounds, for which Gi is of the same order as in YBCO \cite{jan}. Nevertheless, as we will see below, there is a qualitative agreement between the behavior of  $H_{c2}^c$ estimated for polycrystalline material, and the behavior of $H_{c2}^c$ measured in single crystals where it can be defined as a true onset of superconductivity at 90\% or 99\% of the normal state resistivity for the corresponding field orientation.

Figure 2 shows the respective phase diagrams for the different
doping levels displayed in Fig. 1. Dark blue symbols correspond to
the onset of the superconducting transition (corresponding to 90 \% of the
resistivity of the normal state just prior to the
resistive transition), blue markers depict the midpoint of the transition,
while red ones correspond to the foot the superconducting
transition defined as $\rho(T,H)=0.1\rho_n$. Green markers indicate
the position in field and temperature where the resistivity of our
samples drops below our experimental sensitivity. These points could
reflect the behavior of the irreversibility field if one could
neglect inter- or intra-granularity effects which are relevant
for polycrystalline material. As we will see below, this
``irreversibility field" obtained in this way behaves very differently from
the melting field $H_m$ measured in single crystals. Furthermore, as is clearly seen
in Fig. 2, these points move very quickly to low temperatures as
small fields are applied, which would either suggest an extreme
two-dimensionality (which contradicts the observed anisotropy in
$H_{c2}$ and in the $\rho$ for these materials) or strong granularity
effects.

Two important aspects can be clearly seen from the data: i) the
extrapolation of the upper critical fields to zero temperature
$H_{c2}(0)$ estimated either from the slope of $H_{c2}(T)|_{T
\rightarrow T_c}$ through the Werthamer-Helfand-Hohenberg
expression or through the simple phenomenological expression
$H_{c2}(T) = H_{c2}(0)(1-(T/T_c)^2)$ yields $H_{c2}(0)$ values which
are several times larger than $k_BT_c / \mu_B$, particularly for
in-plane fields (for example, for $x = 0.2$ one has $T_c \simeq 48$K
while $H_{c2}^{ab}(0)$ obtained for the onset of the SC transition
from both methods yields respectively, $H_{c2}^{ab}(0) = (247 \pm
17)$ T and $(221 \pm 15)$ T), and ii) the position of the foot of the
resistive transition which is expected to represent the
upper critical field $H_{c2}^c(T)$ along the c-axis shows an upward curvature or a
nearly divergent behavior as the temperature is lowered. Observation
i) may be indicative of unconventional superconductivity and perhaps
strong-coupling pairing or even for a prominent role for the spin-orbit interaction, since the values for $H_{c2}(0)$
quoted above are several times larger than the the weak-coupling Pauli limiting
field, $H_p [T] \simeq 1.84 k_B T_c[K]$ . The observation ii) has
been interpreted in terms of a two-gap superconducting pairing
\cite{hunte}, in line with the extended s-wave model \cite{mazin}.
As was first pointed out by Ref. \cite{jan2},
polycrystalline 1111 materials exhibit an appropriate scaling, which
enable us to collapse all the graphs in Fig. 2, i.e. all the superconducting phase
diagrams for the different doping levels, in a single diagram as
shown in Fig. 3 (a). Here the points correspond respectively,
to the onset, middle point and foot of the resistive transitions
under field normalized with respect to the square of their values for
the superconducting transition temperature $(T_c^2)$ at each doping, and as functions of the reduced temperature $t=T/T_c$. This
scaling has a simple explanation based on the small size of the
coherence lengths in oxypnictides.  The in-plane and the inter-plane
upper critical fields, $H_{c2}^{ab} (T) \simeq \phi_0 / 2 \pi \xi_{ab}(T) \xi_c(T)$ and $H_{c2}^{c} (T) \simeq \phi_0 / 2 \pi \xi^2_{ab} (T)$ ($\phi_0$ is the magnetic flux quantum), are
determined by the respective out-of-plane and in-plane coherence
lengths $\xi_{c} = \xi_{ab} \gamma^{-1/2}$ where $\gamma= m_c / m_{ab}$ is the effective mass anisotropy parameter. In a
\emph{clean} limit for which the mean free path $\ell$ is much
greater than $\xi$ at $T << T_c$, the coherence lengths scale like
$T_c^{-1}$, leading naturally to $H_{c2} \propto T_c^2$. In other
words, as argued in Ref. \cite{jan2}, despite the presence of
impurity phases, as discussed in the previous section, and the
granularity inherent to a polycrystalline material, the coherence
lengths in this material are small enough for the Cooper pairs not
to be significantly affected by the impurities, grain size or
inter-grain coupling. There are two main differences between this
graph and a similar one shown in Ref. \cite{jan2} i) it extends this
scaling over a broad range in $t=T/T_c$, where the upward curvature
of $H_{c2}^{c}$ becomes quite evident, and ii)
$H_{c2}^{ab}$ or the onset of the resistive transition
also seems to satisfy the scaling. It further suggests that
$H_{c2}^{ab} (0) ~ \sim 0.13 T_c^2$, which for a maximum $T_c
= 55$ K would lead to $H_{c2}^{ab} (T = 0)\sim 393$ T or
nearly four times the value expected for $H_p$. However,
one should keep in mind that the extrapolation of $H_{c2}$ measured
near $T_c$ to lower temperatures following
Ginzburg-Landau theory, disregards the essential paramagnetic
limitations and could overestimate the actual $H_{c2}(0)$ values \cite{jan}.

Figure 3 (b) shows $H_{c2}^{c}$ (red markers) and $H_{c2}^{ab}$ (black markers)
as extracted from the WHH formula, i.e $H_{c2}(0) \simeq - 0.69 \partial H_{c2}(T)/\partial T|_{T \rightarrow T_c} T_c$
(squares) and from the phenomenological equation $H_{c2}(T) \simeq H_{c2}(0)(1-(T/T_c)^2)$ (circles) as a function of the fluorine doping level $x$. It also shows the anisotropy $\gamma = H_{c2}^{ab}/ H_{c2}^{c}$ obtained from both forms of estimating $H_{c2}(0)$. Both estimations suggest that $\gamma$ increases with F content saturating at a value $\sim 3.5$. It is nevertheless quite unexpected that one is able to find such a scaling relation when the anisotropy is changing with the doping content. It is  actually even more remarkable given the fact that this relation does not hold in double layered single crystals, which one would expect to be cleaner, as we will discuss below.

\subsection{Single-crystals of single-layered compounds}

\begin{figure}[htb]
\begin{center}
\epsfig{file= 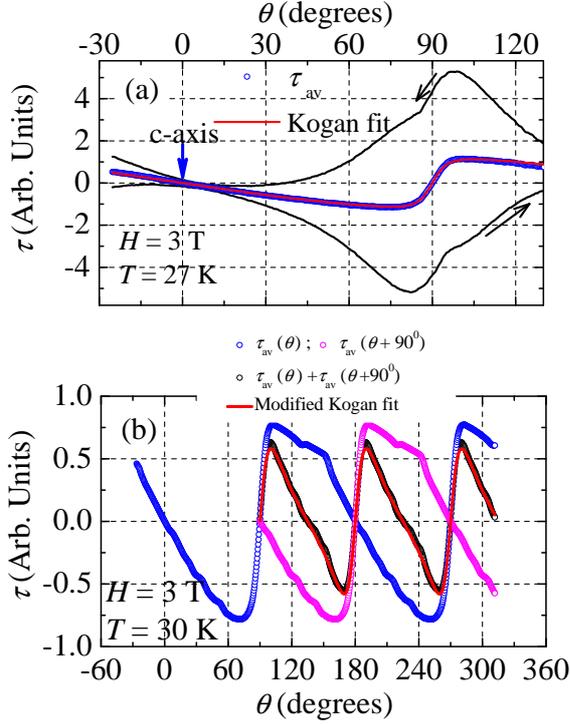, width = 7.4 cm} \caption {(color
online) (a) Magnetic torque $\tau(\theta)$ for a
SmOFeAs$_{0.8}$F$_{0.2}$ single crystal for increasing and
decreasing angle sweeps (black lines) at 3 T and 27 K. The
reversible equilibrium $\tau_{\textrm{\scriptsize{av}}}(\theta)$ (blue markers) is obtained by
averaging both traces. Red line is a fit to the Kogan expression from which one can
extract the anisotropy parameter $\gamma$, in-plane penetration depth $\lambda_{\textrm{\scriptsize{ab}}}$, and inter-plane upper critical field $H_{c2}^{c}$, see text. (b) Angular dependence of $\tau_{\textrm{\scriptsize{av}}}(\theta)$ (blue)
and $\tau_{\textrm{\scriptsize{av}}}(\theta + 90^{\circ})$ (magenta) for 3 T and
30 K. Such a procedure described in Ref. \cite{me} suppresses the magnetic component in the torque associated to Sm or Fe ions. Red line corresponds to a fit of
$\tau_{\textrm{\scriptsize{av}}}(\theta)+ \tau_{\textrm{\scriptsize{av}}}(\theta + 90^{\circ})$
to a modified Kogan expression.}
\end{center}
\end{figure}
\begin{figure}[htb]
\begin{center}
\epsfig{file= 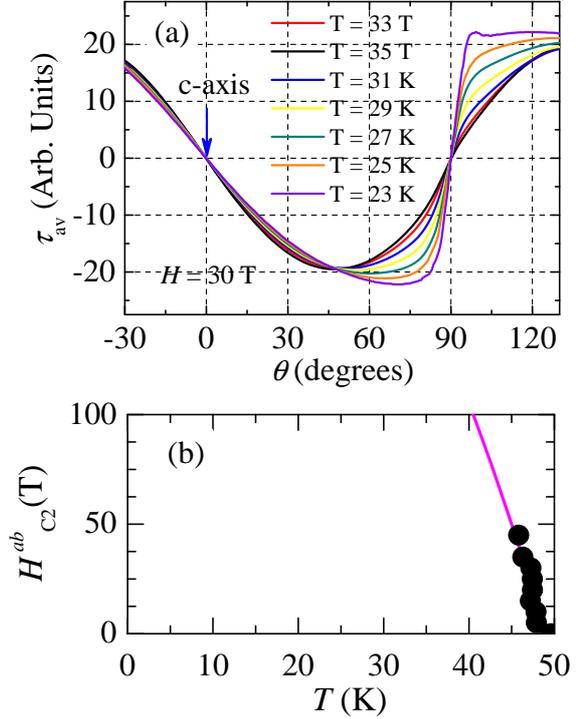, width = 7.4 cm} \caption {(color
online) (a) Angular dependence of $\tau_{\textrm{\scriptsize{av}}}(\theta)$ at 30
T and for several temperatures. (b) $H_{c2}^{ab}$ from the resistive transition in polycrystalline SmFeAsO$_{0.82}$F$_{0.18}$ (black markers) and fit to the expression
$H_{c}^{ab}(T) = H_{c}^{ab}(0) [1-(T/T_c)^2]$ and which was used as input in Eq. 2. For all our fittings $\eta = 1$.}
\end{center}
\end{figure}

The high $T_c$ values and the extremely high upper
critical fields of the oxypnictides \cite{hunte} could
make them promising candidates for power applications if,
unlike the layered cuprates, a sizeable vortex liquid region
responsible for dissipative flux-flow does not dominate a large portion of their
temperature-magnetic field $(T-H)$ phase diagram. Therefore
it is important to study the response of the anisotropic
magnetization in the vortex state of the oxypnictides, in particular,
the extent to which  vortex properties are affected by strong
magnetic correlations, multiband effects and a possible interband
phase shift between order parameters on the different sheets of the
Fermi surface \cite{theory1}. For instance, multiband effects are known
to manifest themselves in MgB$_2$ as a strong temperature and field
dependency of the mass anisotropy parameter $\gamma(T,H)$ and of the
London penetration depth $\lambda(T,H)$ even at $H\ll H_{c2}$
\cite{MgB2vsT, MgB2vsH}.

Measurements of the anisotropic equilibrium magnetization $m(T,H)$ in
 1111 single crystals are complicated by the smallness of
$m(H,T)$ caused by the large Ginzburg-Landau parameter,
$\kappa=\lambda/\xi > 100$ and by the strong paramagnetic response of the
lanthanide and Fe ions, which can mask the true behavior of $m(T,H)$. In
this situation torque magnetometry is the most sensitive technique
to measure fundamental anisotropic parameters in $\bf{m}(T,H)$
particularly in small single crystals. The torque $\tau={\bf m}\times{\bf H}$
acting upon a uniaxial superconductor is given by \cite{kogan, koganprl}
    \begin{eqnarray}
    \tau(\theta) = \frac{HV \phi_0 (\gamma^2-1)\sin2\theta}{16\pi \mu_0 \lambda_{ab}^2\gamma\varepsilon(\theta)}
    \ln \left[ \frac{\eta H_{c2}^{ab}}{\varepsilon(\theta) H}\right]
    \nonumber \\
    + \tau_m\sin 2\theta,
    \label{Kogan}
    \end{eqnarray}
where $V$ is the sample volume, $\phi_0$ is the flux quantum,
$H_{c2}^{ab}$ is the upper critical field along the ab planes,
$\eta\sim 1$ accounts for the structure of the vortex core,
$\theta$ is the angle between $\bf{H}$ and the c-axis,
$\varepsilon(\theta) = (\sin^2 \theta+\gamma^2\cos^2\theta)^{1/2}$
and $ \gamma = \lambda_c/ \lambda_{ab}$ is the ratio of the London
penetration depths along the c-axis and the ab-plane. The first
term in Eq. (\ref{Kogan}) was derived by Kogan in the London
approximation valid at $H_{c1}\ll H\ll H_{c2}$ \cite{kogan}. The
last term in Eq. (1) describes the torque due to the paramagnetism of
the lanthanide oxide layers and possible intrinsic magnetism of the FeAs
layers. Here, $\tau_m=(\chi_c-\chi_a)VH^2/2$ and $\chi_c$ and
$\chi_a$ are the normal state magnetic susceptibilities of a
uniaxial crystal along the c axis and the ab plane, respectively.

As we showed recently \cite{me}, the paramagnetic term in Eq. (\ref{Kogan}) in
SmO$_{0.8}$F$_{0.2}$FeAs single crystals with $T_c \sim 45$ K can be larger
than the superconducting
torque, which makes extraction of the  equilibrium vortex magnetization rather nontrivial.
This problem was circumvented by applying a simple mathematical procedure, which
enabled us to unambiguously extract the superconducting component of the torque from the
data by fitting the sum
$\tau_{\textrm{\footnotesize{av}}}(\theta)+\tau_{\textrm{\footnotesize{av}}}(\theta + 90^{\circ})$,
in which the odd harmonics due to the paramagnetic component cancel out:
    \begin{eqnarray}\label{Koga}
    \tau(\theta)+ \tau(\theta + 90^{\circ}) = \frac{V \phi_0(\gamma^2-1) H\sin2 \theta}{16\pi \mu_0 \lambda_{ab}^2\gamma}
    \nonumber \\
    \times \left[ \frac{1}{\varepsilon(\theta)} \ln \left(
    \frac{\eta H_{c2}^{\|}}{\varepsilon(\theta) H}
    \right) - \frac{1}{\varepsilon^{\star}(\theta)} \ln \left( \frac{\eta H_{c2}^{\|}}{\varepsilon^{\star}(\theta) H}
    \right) \right],
    \end{eqnarray}
where $\varepsilon^{\star}(\theta) = ( \cos^2\theta + \gamma^2
\sin^2\theta )^{1/2}$ and $\eta \simeq 1$.

The above discussion is illustrated by Figure 4. For instance, Fig. 4 (a) shows $\tau(\theta)$ for a SmO$_{0.8}$F$_{0.2}$FeAs single crystal in a limited angular range and for both
increasing and decreasing angular sweeps at $T= 27$ K and $H = 3$ T. The reversible component $\tau_{av}(\theta)$
is extracted by taking the average of both ascending and descending branches of
$\tau(\theta)$. This procedure gives a characteristic behavior
 of $\tau_{av}(\theta)$ almost perfectly described by Eq. (\ref{Kogan}) represented by the red line.
The fit yields a value $\gamma \simeq 11.5$
and a value for $H_{c2}^{ab}= \gamma H_{c2}^{c}$ that is considerably lower than the value extracted from our polycrystalline samples having a similar $T_c$ (shown in Fig. 5 (b)). The reason is that the direct 3 parameter fit using Eq. (\ref{Kogan}) is not very stable, allowing many different fit parameters to give equally good description
of the torque data.

As mentioned above, the paramagnetic torque component $\tau_m\propto H^2$ becomes particularly pronounced
at very high fields for which it can become larger than the equilibrium superconducting torque $\tau_s\propto H\mbox{ln}(H_{c2}/H)$. As a result, the application of the procedure based on Eq. (\ref{Koga}) becomes the only way to  unambiguously extract the relatively small component $\tau_s$ from the torque data.  This is illustrated by Fig. 4 (b) which shows $\tau_{\textrm{\footnotesize{av}}}(\theta)$, $\tau_{\textrm{\footnotesize{av}}}(\theta + 90^{\circ})$ and $\tau_{\textrm{\footnotesize{av}}}(\theta)+ \tau_{\textrm{\footnotesize{av}}}(\theta + 90^{\circ})$, as a function of $\theta$ for $H = 3 $ T and $T = 30$ K. As follows from Fig. 4(b), the component of the torque due solely to the superconducting response shows a sharper and less ``rounded" angular dependence. The important aspect is that this procedure not only gives a smaller value for $\gamma = 8.7$ but it also leads to an $H_{c2}^{c}$ value, which is more consistent with our transport measurements in polycrystalline material. Although a word of caution is needed here. In multi-band superconductors, as is possibly the case for the oxypnictide superconductors, the superconducting anisotropy as extracted from the ratio of penetration depths may differ considerably from the anisotropy as extracted from the ratio of the upper critical fields. The Kogan formalism was developed assuming a single band scenario although it was latter extended to include possible differences among both anisotropies \cite{koganprl}. Given the lack of a proper multi-band formalism to analyze our torque data, our current strategy is to extract $H_{c2}^{c}$ and $H_{c2}^{ab}$ from transport measurements in our single crystals, introduce both $H_{c2}^{c}$ and the anisotropy ratio in $H_{c2}$ within the extended Kogan formula \cite{koganprl} re-fitting our original data to extract the anisotropy in penetration depth, after subtraction of the paramagnetic component. Following our work, a similar strategy was already applied by Ref. \cite{weyeneth2}. This would avoid discrepancies in $H_{c2}^{c}$ extracted from the transport and the torque data and among the values of $H_{c2}^{c}$ extracted from different torque data sets, these \cite{me} and those of Ref. \cite{weyeneth}. The ability to perform transport measurements in these single crystals was developed only very recently. We expect, nevertheless that our main broad conclusions concerning the qualitative dependence of the anisotropy in temperature and field will remain valid.

Figure 5 (a) shows a data set of $\tau_{\textrm{\footnotesize{av}}}(\theta)$ taken under $H = 30$ T and for several temperatures, for which we applied the procedure illustrated by Fig. 4 (b).
As discussed in Ref. \cite{me}, our systematic study of $\tau_{\textrm{\footnotesize{av}}}(\theta)$ gives $\gamma(H, T)$, which exhibits the field and temperature dependencies similar to those previously seen in MgB$_2$, further supporting  the two-gap scenario for the single layered oxypnictides. For fields all the way up to 30 T, we could not detect any noticeable effect of the field on the in-plane penetration depth (or equivalently on the superfluid density) that would indicate the suppression of a superconducting gap in one of the bands. This is remarkable, since as seen in Fig. 5 (b) $30$ T is very close to the values of $H_{c2}^{c}$ (open markers), extracted from two sets of torque data measured at respectively 3 and 30 T and in the temperature range between 25 and 30 K, and suggests strong coupling superconductivity (perhaps two very strongly coupled gaps), or perhaps even an abrupt first order transition \cite{lev}.

\subsubsection{Evaluation of the irreversibility line.}
\begin{figure}[htb]
\begin{center}
\epsfig{file= 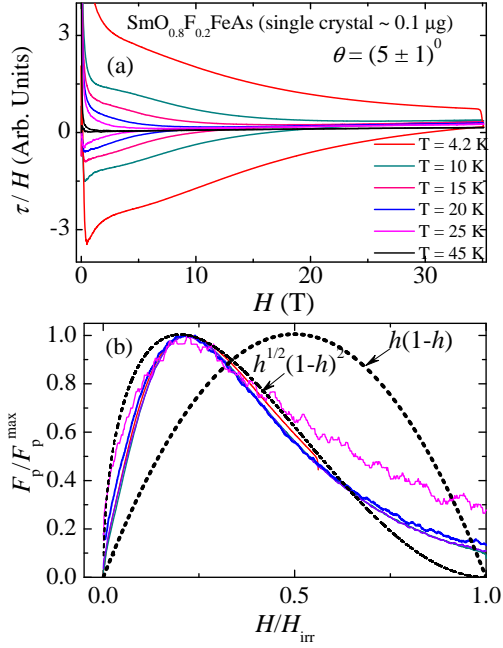, width = 6.5 cm} \caption {(color
online) (a) Magnetic torque $\tau$ normalized by the field $H$ and as a function field for a SmOFeAs$_{0.8}$F$_{0.2}$ single crystal and for several temperatures. The vortex pinning force $F_p$ is proportional the irreversible component in the torque, i.e., $\Delta \tau /H = H^{-1}[\tau(H_{\uparrow}) - \tau(H_{\downarrow})$. (b) Pinning force $F_p$ normalized with respect to its maximum value $F_p^{\textrm{\scriptsize{max}}}$ as a function of the reduced field $h = H/H_{\textrm{\scriptsize{irr}}}$. }
\end{center}
\end{figure}
\begin{figure}[htb]
\begin{center}
\epsfig{file= 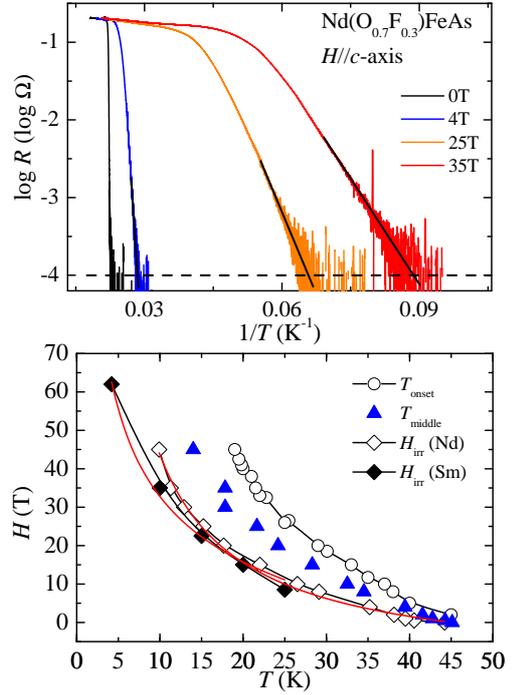, width = 6.5 cm} \caption {(color
online) Top panel: Logarithmic dependence of the resistive transition in a NdOFeAs$_{0.7}$F$_{0.3}$ single crystal as function of the inverse temperature $T^{-1}$, from Ref. \cite{jan}. The temperature at which the resistivity reaches the level of noise of our experimental set-up may reflect the behavior of the irreversibility line. Bottom panel: An attempt to draw the irreversibility line in the superconducting phase diagram of the 1111 compounds from both transport and torque measurements. Circles and rectangles define respectively the onset and the middle point of the resistive transition in a NdOFeAs$_{0.7}$F$_{0.3}$ single crystal. Open rectangles depict the ``irreversibility line" from transport measurements, while solid rectangles depict the irreversibility field extracted from our scaling analysis in Fig. 6 for SmOFeAs$_{0.8}$F$_{0.2}$. Red lines are fits to the expression $H_m (t) \propto t^{-\alpha}(1-t)$ with $1 < \alpha < 2$ (see text).}
\end{center}
\end{figure}

 Using the torque magnetometry technique it is also possible to extract the irreversibility field $H_m(T)$ by sweeping the field up and down at a given temperature and measuring the hysteretic magnetization loop, $m=\Delta \tau (H)/ H = H^{-1}(\tau (H_{\uparrow})+ \tau (H_{\downarrow}))/2$. The field at which the magnetization loop closes defines the irreversibility line $H_m(T)$ that separates pinned vortex vortex solid and vortex liquid states. Below the irreversibility field the torque measurement of the irreversible component of the magnetization $\Delta m(H)=\Delta \tau (H)/H$ enable us to  use the Bean critical state model \cite{bean} to extract the critical current density of a superconductor, $J_c = k \Delta m(H)/R$, and the pinning force $\textbf{F}_p(H) = \textbf{J}_c(H) \times \textbf{H}$ per unit volume where $k$ is a constant that depends on the geometry of the sample and $R$ is the dimension of the sample perpendicular to the field direction.  Thus, we have $J_c (H)\propto \Delta \tau(H)/H$ and $F_p (H)\propto \Delta \tau(H) $.
For conventional type-II superconductors, scaling laws for flux pinning proposed by Dew-Hughes, \cite{hughes} Campbell and Evetts, \cite{evetts} or by Kramer \cite{kramer} can be written in a general form as $F \propto H_{c2}^m h^p(1-h)^q$, with $h=H/H_{c2}$.  For high $T_c$-cuprates superconductors, similar phenomenological scaling relations have been used \cite{yamasaki}, but with a different definition of the reduced field,  $h=H/H_{\textrm{\footnotesize{irr}}}$ so that the critical current vanishes at $H_{\textrm{\footnotesize{irr}}}$.

Figure 6 (a) shows raw torque normalized by the field data for a SmOFeAs$_{0.8}$F$_{0.2}$ single crystal as a function of field $H$ applied nearly along the c-axis, or more precisely for an angle $\theta \simeq 5 \pm 1 ^{\circ}$ between  $H$ and the c-axis of the crystal as measured via a Hall probe. A very strong hysteresis, i.e., the irreversible component in the $\tau (H)/H$, emerges as the temperature is lowered, from which we can estimate the overall behavior of $F_p$.
In the cuprates the pinning force curves taken at different temperatures often collapse into a scaling relation  $F_p/F_p^{\textrm {\footnotesize{max}}}\propto h^p(1-h)^q$ where $F_p^{\textrm {\footnotesize{max}}}$ is the maximum pinning force, and  $h=H/H_{\textrm{\footnotesize{irr}}}$ \cite{pin}. As follows from Fig. 6 (b), the resulting pinning curves for the oxypnictides also tend to collapse into an asymptotic behavior given by the expression $F_p/F_p^{\textrm {\footnotesize{max}}}\propto h^{1/2} (1-h)^2$, similar to that of the Kramer model \cite{kramer} in which $J_c$ is determined by shear depinning of the vortex lattice. Figure 6 (b) also shows the asymptotic behavior $F_p/F_p^{\textrm {\footnotesize{max}}}\propto h(1-h)$ characteristic of a dense array of strong core pinning centers in NbTi \cite{nbti}. It is clear that $F_p/F_p^{\textrm {\footnotesize{max}}}\propto h^{1/2} (1-h)^2$  provides a much better description of our data, indicating that the field dependence of the pinning force in our SmOFeAs$_{0.8}$F$_{0.2}$ sample is more reminiscent of that of Nb$_3$Sn \cite{nb3sn} and YBa$_2$Cu$_3$O$_{7-x}$ \cite{pin}.

The irreversibility field $H_{\textrm{\footnotesize{irr}}}$ extracted from the analysis described above exhibits the temperature dependence $H_{\textrm{\footnotesize{irr}}}(t) \propto (1-t)/t^\alpha$ with $\alpha \approx 1$ for  NdOFeAs$_{0.7}$F$_{0.3}$, and
$\alpha\approx 0.6$ for SmOFeAs$_{0.8}$F$_{0.2}$ (see Fig. 7b). Such behavior of $H_{\textrm{\footnotesize{irr}}}(T)$ is similar to that of the melting field of the vortex lattice in moderately anisotropic uniaxial superconductors, for which $H_m (t) \propto t^{-\alpha}(1-t)^\beta$ with $1 < \alpha < 2$, $1 < \beta < 2$ \cite{blatter}. This temperature dependence along with the fact that $H_{\textrm{\footnotesize{irr}}}$ lies noticeably below $H_{c2}$ indicate that vortex fluctuations in 1111 oxypnictides are rather strong, consistent with  the estimations of the Ginzburg number $Gi$, which characterizes fluctuations of the order parameter near $T_c$. As shown earlier \cite{jan,jan2}, the value of $Gi\sim 10^{-2}$ for NdOFeAs$_{0.7}$F$_{0.3}$ and SmOFeAs$_{0.8}$F$_{0.2}$ is of the same order of magnitude as $Gi$ for YBa$_2$Cu$_3$O$_{7-x}$.

Significant vortex fluctuations in NdOFeAs$_{0.7}$F$_{0.3}$ single crystals also manifest themselves in the clear Arrhenius plot of the resistance shown in Fig. 7(a). This data that was already reported in Ref. \cite{jan}, provide an unambiguous evidence for thermally-activated vortex dynamics in 1111 oxypnictides in a wide range of temperatures and fields. Our goal here is to evaluate $H_{\textrm{\footnotesize{irr}}}$ from these transport measurements in order to compare it with $H_{\textrm{\footnotesize{irr}}}$ extracted from our torque measurements in SmOFeAs$_{0.8}$F$_{0.2}$. For NdOFeAs$_{0.7}$F$_{0.3}$, which has a $T_c$ close to the value reported for our SmOFeAs$_{0.8}$F$_{0.2}$ single crystal, we define the irreversibility field at which the resistivity drops below our experimental sensitivity (horizontal line in Fig. 7 (a)). As follows from Fig. 7 (b), the agreement between our crude estimate from the transport data and the $H_{\textrm{\footnotesize{irr}}}$ extracted from the torque data is quite remarkable, except perhaps for the points from the transport data at the highest fields which are subjected to a lower signal to noise ratio.

\subsection{Single-crystals of double-layered compounds}

In this section we present a detailed high field magneto-transport study in the so-called double layered or 122 Fe arsenide compounds with the goal of extracting their phase diagram and drawing a comparison with the previously presented data on the 1111 compounds by addressing the following questions. Is the upward temperature dependence of $H_{c2}(T)$ along the c-axis a general feature of the Fe arsenide compounds? To what extent does a magnetic field broaden the phase diagram of the 122 compounds as compared to that of the 1111s, or in other words, how similar is the vortex physics of the 122s with respect of that of the 1111s? Are the upper critical fields of these compounds at low temperatures as tremendously enhanced with respect to the weak coupling Pauli limiting field as those of the 1111 compounds? Is there any possibility of observing new superconducting phases, such as the Fulde-Ferrel-Larkin-Ovchinnikov (FFLO) state \cite{fflo} in these compounds? The search for answers to these questions was the motivating factor behind our preliminary exploration of the 122 compounds.
\begin{figure}[htb]
\begin{center}
\epsfig{file= 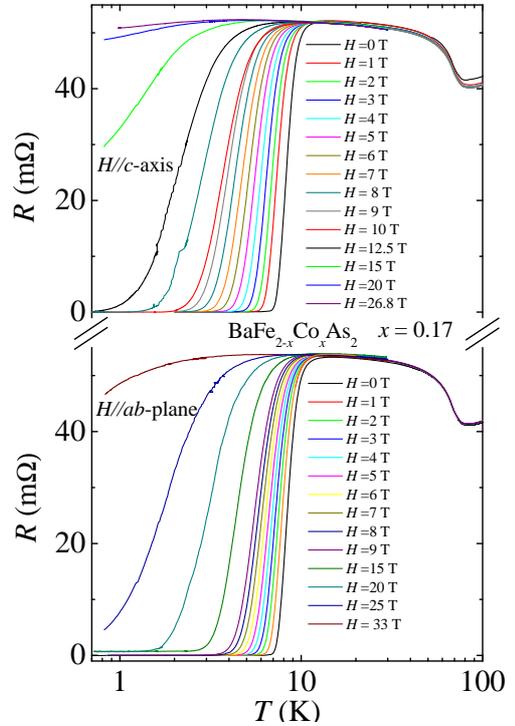, width = 6.5 cm} \caption {(color
online) (a) Resistance $R$ as a function of temperature for a BaFe$_{2-x}$Co$_x$As$_2$ single crystal having a nominal doping of 0.17 and for several field values oriented along
an in-plane direction. (b) Same as in (a) but for fields along the inter-plane direction. The step observed at $T \sim 60$ K corresponds to the transition towards the antiferromagnetic state \cite{chen2}.}
\end{center}
\end{figure}
\begin{figure}[htb]
\begin{center}
\epsfig{file= 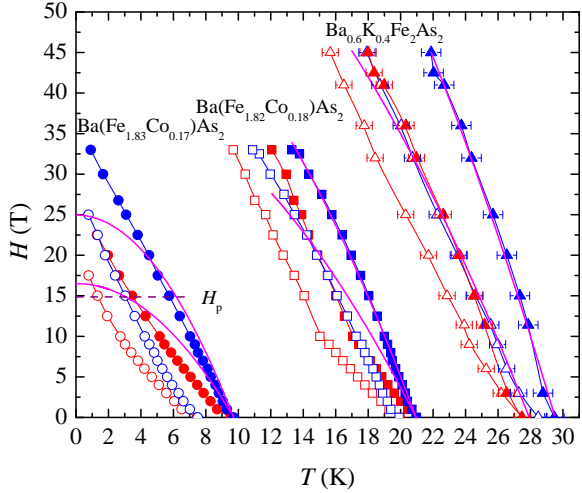, width = 6.5 cm, angle=-90} \caption {(color
online) Magnetic field $H$ temperature $T$ superconducting phase diagram of several underdoped double-layered Fe arsenide compounds. Blue and red symbols depict the phase
boundary for fields along an in-plane and the inter-plane directions, respectively. Closed and open symbols depict respectively the onset (90 \%) and the foot or 10 \% of the resistive transition. Triangles, squares and circles, represent the boundaries for  Ba$_{0.6}$K$_{0.4}$Fe$_2$As$_2$, BaFe$_{1.82}$Co$_{0.18}$As$_2$, and BaFe$_{1.83}$Co$_{0.17}$As$_2$,  respectively. Magenta lines are an attempt to the fit in each case the onset of the resistive transition to $H_{c2}(T)=H_{c2}(0)[1-(T/T_c)^2]$. For Ba$_{0.6}$K$_{0.4}$Fe$_2$As$_2$ we simultaneously measured two single-crystals with slightly different $T_c$s, one with the field oriented along the ab-plane and the other with the field applied along the c-axis.}
\end{center}
\end{figure}
Figure 8 (a) and (b) show a typical set of resistance data for a BaFe$_{1-x}$Co$_x$As$_2$ single crystal (nominal doping $x=0.17$) as a function of temperature for several field values and two field orientations along the c-axis and along the ab plane, respectively. Here we show only data for a few samples whose superconducting transition width at zero field $\Delta T_c (H = 0) \lesssim 2$ K. In marked contrast with the 1111 compounds the width of superconducting transition is modestly affected by the external field, see for example similar data on Refs. \cite{tillman, yamamoto}. Thus, the behavior of 122 compounds under field is much more akin to that of the low $T_c$ superconducting compounds where the resistive transition does not broaden much but shifts to lower temperatures as a larger field is applied.

From several data sets like this we can come up with a superconducting phase diagram for several 122 compounds shown in Fig. 9. It contains data for two nominal concentrations
of BaFe$_{2-x}$Co$_x$As$_2$, namely $x = 0.17$ and 0.18, and Ba$_{1-x}$K$_x$Fe$_2$As$_2$ with a nominal concentration $x = 0.4$. The middle point of the resistive superconducting
transition at zero field for each compound takes place at, $T = (8.1 \pm 0.1)$ K, $(20 \pm 0.05)$ K, and  $(28 \pm 0.1)$ K, respectively. In Fig. 9 blue and red markers depict the
superconducting phase boundary respectively for $H$ parallel and perpendicular to the FeAs planes. Closed and open symbols depict respectively the onset (90 \%)
and the foot or 10 \% of the normal-state resistivity. 

The upward curvature seen in $H_{c2}$ for Ba$_{0.6}$K$_{0.4}$Fe$_2$As$_2$ and for temperatures close to $T_c$ may result from  a distribution of superconducting transition temperatures, possibly due to a spread in the local doping levels. In this case our transport measurements of Hc2 defined at 90\% of the normal state resistivity mostly probe the superconducting properties of the regions with optimal doping for which $T_c$ is maximum. It is interesting to compare our data with the heat capacity measurements of $H_{c2}(T)|_{T \rightarrow T_c}$ on the same material, which reveals a linear temperature dependence for $H_{c2}(T)$ \cite{welp}. Since the heat capacity measurements probe $H_{c2}(T)$  averaged  over the entire  distribution of $T_c$s, the difference between the transport and the specific heat measurements may reflect the difference in the temperature dependencies of $H_{c2}(T)$ in optimally doped and non optimally doped regions.

Several important observations follow from the data shown in Fig. 9: i) $H_{c2}(T)$ as a function of temperature for $H \parallel$ c is also concave as it is for the 1111 compounds, ii)
while $H_{c2}(T)$ for $H \bot$ c behaves nearly linearly in temperature, at least as seen for the BaFe$_{1.83}$Co$_{0.17}$As$_2$ compound for which we were able
to explore a larger portion of its superconducting phase diagram, iii) the upper critical fields extrapolated to zero temperature for both field orientations are clearly well
beyond the week coupling Pauli limiting field, i.e. for BaFe$_{1.83}$Co$_{0.17}$As$_2$ one obtains $H_p = 1.84 \times 8.1 \simeq 14.9 $ T. In reality, for fields applied along the ab-plane
the extrapolation of $H_{c2}$ to zero temperature gives values about twice the BCS paramagnetic limit.  Notice, that for reasons that are currently unclear, the $H_{c2}(T)$ values found by us for  BaFe$_{2-x}$Co$_x$As$_2$ are considerably larger than the ones reported in Ref. \cite{tillman} for samples having approximately the same $T_c$s.
Our observations immediately suggest a parallelism with strongly coupled superconductors such as the heavy fermion compounds. For example, in the so-called 115 compounds the
ratio $2 \Delta_0 / k_B T_c$ is claimed to be much larger than the BCS value 1.76 \cite{onuki} while $H_{c2}(T)$ along an in-plane direction extrapolated to zero temperature surpasses by
far the expected weak coupling Pauli limiting field \cite{murphy}. It has also been argued, at least for the Co doped compounds, that there might be a finite although small spread in the local
Co concentrations in different parts of the sample \cite{tillman}, thus creating an intrinsic disorder.  Disorder could also contribute to the increase in the upper critical
field in these materials \cite{lev2,jan}. In fact, the phase diagrams of the 122 compounds shown in Fig. 9, do not follow the scaling shown in Fig. 1 for the 1111 compounds, indicating that these nominally single-phase single-crystals may not be in the clean limit. It is interesting to mention, for
instance, that a simple fit of the onset of $H_{c2}(T)$ for Ba$_{0.6}$K$_{0.4}$Fe$_2$As$_2$ for fields along the planar direction to the expression $H_{c2}(T)= H_{c2}(0)(1-(T/T_c)^2)$ (magenta lines in Fig. 9) yields $H_{c2}(0) = (100 \pm 2) $ T, which is a factor of 2 higher than $H_p$. Although, as we have learned from BaFe$_{1.83}$Co$_{0.17}$As$_2$ this expression may strongly underestimate
$H_{c2}(0)$. In all three cases, the anisotropy $\gamma = H_{c2}^{ab}(T)/H_{c2}^{c}(T)$ is observed to decrease from a value $\gamma \gtrsim 3$  for $ T \rightarrow T_c$
to a value $\gamma \gtrsim 1$  for $T << T_c$  as already reported in Ref. \cite{singleton}.

\section{Conclusions}

In conclusion in all oxypnictide compounds explored by us we observe:
\begin{enumerate}
  \item A pronounced upward curvature in the temperature dependencies of  $H_{c2}(T)$ along the c-axis.
  \item Relatively modest effective mass anisotropies which are temperature dependent (as seen in MgB$_2$),
  reaching values in the order of 9 at low temperatures for the 1111 compounds or values ranging between 3 and 1 (at low temperatures and high fields) for the 122 compounds
  \item Upper critical fields extrapolated to zero temperature that are a factor of two to three higher than the weak coupling Pauli limiting field as in the heavy-fermion compounds.
\end{enumerate}

All points listed above are clear indications for unconventional superconductivity in the Fe arsenides: points 1 and 2 are consistent with a two-gap scenario \cite{hunte,jan}, and point 3 may indicate a significant enhancement of the Pauli limiting field by strong coupling effects.
The more anisotropic single-layered 1111 compounds, which display the highest $T_c$s in the oxypnictide family, exhibit a field-temperature superconducting phase diagram similar to that of the least anisotropic cuprates. This manifests itself in the existence of a irreversibility field $H_{\textrm{\footnotesize{irr}}}$ well below $H_{c2}(T)$, ohmic thermally-activated flux flow resistivity at $H_{\textrm{\footnotesize{irr}}}(T)<H<H_{c2}(T)$, and a temperature dependence for $H_{\textrm{\footnotesize{irr}}}$ which is consistent with that of the melting field of the vortex lattice in a moderately anisotropic uniaxial superconductor. Yet despite the relatively strong vortex fluctuations in the 1111 compounds, their phase diagrams do not exhibit the very wide vortex liquid phase regions characteristic of layered Bi-2223 and Bi-2212 cuprates. In fact, as follows from Fig. 9, the irreversibility field $H_{\textrm{\footnotesize{irr}}}$ above which 1111 compounds can carry weakly dissipative currents exceeds 30T at 10K, which is higher than $H_{\textrm{\footnotesize{irr}}}(T)$ for dirty MgB$_2$. These features of the oxypnictides (even in their present and far from optimized condition) along with their extremely high $H_{c2}$ values can make these materials new promising contenders for high-field power applications.



\end{document}